# Twitter and disability activism: leadership and relevant topics in the online conversation

# Twitter y el activismo por la discapacidad: liderazgo y temas relevantes en la conversación en red

# Twitter e ativismo por deficiência: liderança e questões relevantes na conversa online

Terese Mendiguren Galdospin[1]*
Koldobika Meso Ayerdi[1]**
Jesús Ángel Pérez Dasilva[1]***
María Ganzabal Learreta[1]****
Ainara Larrondo Ureta[1]*****
Simón Peña Fernández[1]******

[1] University of the Basque Country (UPV/EHU), Spain

* Associate Professor at the Journalism Department
** Member of the 'Gureiker' Research Group of the Basque University System
*** Senior Lecturer at the Journalism Department
**** Lecturer at the Journalism Department
***** Senior Lecturer in Journalism
****** Lecturer

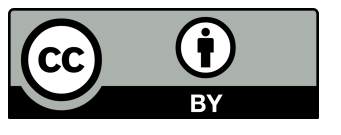

## Abstract

The dissemination and viralization of information on social media has been widely studied from various perspectives, including that of digital activism. On the other hand, disability-related activism has conquered the online environment, thus obtaining a reach that goes beyond the offline space and generating dialogue in the digital sphere. This article analyses the conversation generated on Twitter, taking as a sample all the tweets with the #disability hashtag before and after the International Day of Persons with Disabilities. More than 18,000 tweets, containing almost as many mentions, were analysed and interpreted as the weighted edges of a graph created using Gephi software and applying the Force Atlas 2 brute force algorithm. The focus was placed on the conversational communities generated around that hashtag, their main themes and the prominent participants in them. In conclusion, although the network of mentions is very dispersed, Twitter is the setting for assertions that receive certain institutional and political support in the Latin American environment, for organisations related to the cause (such as the ONCE Foundation) and, above all, the clear predominance of female activists.

## Resumen

La difusión y viralización de la información en los medios sociales ha sido ampliamente estudiada desde diversas perspectivas, incluida la del activismo digital. Por otro lado, el activismo relacionado con la discapacidad ha conquistado el terreno online obteniendo así un alcance que va más allá del espacio offline y generando diálogo en la esfera digital. Este artículo analiza la conversación generada en Twitter tomando como muestra todos los tuits con el hashtag #discapacidad anteriores y posteriores al Día Internacional de las Personas con Discapacidad. Se han analizado más de 18.000 tuits en los que hay casi otras tantas menciones, que han sido interpretadas como las aristas ponderadas de un grafo dirigido con el software Gephi, aplicando el algoritmo de fuerza bruta Force Atlas 2. Se ha puesto el foco en las comunidades conversacionales generadas en torno a ese hashtag, las temáticas principales y los actores destacados en dichas comunidades conversacionales. Se concluye que, aunque la red de menciones es muy dispersa, Twitter es escenario de reivindicaciones que abarcan cierto apoyo institucional y político del entorno latinoamericano, de entidades relacionadas con la causa como La Fundación ONCE y, sobre todo, la predominancia clara de activistas mujeres.

**Resumo**

A disseminação e viralização de informações nas redes sociais tem sido amplamente estudada sob diversas perspectivas, inclusive a do ativismo digital. Por outro lado, o ativismo relacionado à deficiência conquistou o terreno online, obtendo assim um alcance que vai além do espaço offline e gerando diálogo na esfera digital. Este artigo analisa a conversa gerada no Twitter tomando como amostra todos os tweets com a hashtag #deficiência antes e depois do Dia Internacional da Pessoa com Deficiência. Foram analisados mais de 18.000 tweets com quase o mesmo número de menções, que foram interpretados como arestas ponderadas de um gráfico direcionado com o software Gephi, aplicando o algoritmo de força bruta Force Atlas 2. O foco foi colocado nas comunidades de conversação geradas em torno dessa hashtag, os principais temas e os atores de destaque nas referidas comunidades de conversação. Verifica-se que, embora a rede de menções seja muito dispersa, o Twitter é palco de reivindicações que incluem certo apoio institucional e político do ambiente latino-americano, de entidades ligadas à causa como a Fundación ONCE e, sobretudo, o claro predominância de mulheres ativistas.

# 1. Introduction

In Spain, there are over 4.3 million people with some type of disability, according to the Survey on Disability, Personal Autonomy and Dependency Situations produced by the Spanish Statistics Institute (INE, 2022). This collective often faces physical or cognitive barriers that hinder their communication with others and their ability to live full lives in society. These communication barriers are especially suffered by those with hearing, speaking, reading, writing or comprehension problems.

Disabled people are very aware that their disabilities are perceived as a stigma and are detrimental in social environments (Goffman, 1963). Moreover, a negative image has even been constantly transmitted in the media throughout their lives (Oliver, 2004).

In this context, the emergence of social media has not only opened up extensive opportunities for disabled people (Rugoho, 2020), but has also allowed them to control how and when to disclose information about their disability (Furr et al., 2016). It offers new opportunities to forge relationships and achieve social support (Suriá, 2017), which contributes to increased integration of those who suffer the greatest isolation outside the online environment (Banjanin et al., 2015). Social media provides disabled people with new tools for presence, communication and empowerment (Gelfgren et al., 2022). Narratives based on empowerment, humour or responsibility, amongst other factors, are precisely what allow disabled influencers to present themselves on social media not as victims or superhuman figures, but as complex human beings (Södergren y Vallström, 2022).

## 2. Online Disability Activism

Social media, especially YouTube, Instagram and TikTok, have become the ideal location for activism for disabled people (Lapierre, 2023), as these platforms allow for self-representation of disability by content creators themselves (Bonilla del Río et al., 2022b). A fundamental form of activism on social media is storytelling. Marginalised groups are offered alternative means for telling stories, which can complement, complicate and criticise the dominant media narrative (Liao, 2019). In the specific case of disabled activists, storytelling allows them to make their voices heard (Bitman, 2023).

As covered by Bassey et al. (2021), social media boosts this collective's social contact and promotes defence and awareness of disability, in addition to communication, exchange and activism (Ellis and Kent, 2016). In fact, multiple authors present digital activism as an opportunity for digital inclusion of disabled people (Ellis & Goggin, 2018; Bonilla del Río et al., 2021).

They highlight research centred around digital activism aimed at protecting the rights of individuals with functional diversity (Auxier et al., 2019), improving accessibility in ITC use (Tjokrodinata et al., 2022), challenging government policy and negative stereotypes (Pearson & Trevisan, 2015) and promoting political and citizen participation (Mann, 2018; Trevisan, 2020), amongst other topics. Moreover, 'there is a notable change in the social discourse defending diversity in beauty and fashion' (Mañas et al., 2022, p. 206).

In January 2017, the Disability March was created: a social media campaign that allowed activists to protest in an accessible way whilst also presenting and constructing their identities (Li et al., 2018). This initiative showed how, for the first time, disability challenged the relationship between offline and online activism (Dube, 2020). It gradually became clear that revealing a previously-hidden disability contributed to individuals' identities being legitimate and authentic (Evans, 2019).

Multiple studies have shown that social media plays a crucial role in promoting the visibility and acceptance of disability, which contributes to recognition of diversity and social inclusion (Foster and Pettinicchio, 2021). These platforms make disabled people feel less alone (Caron and Light, 2016) and lead them to develop and maintain other meaningful relationships, increase the frequency and quality of their social interactions and reduce their feelings of loneliness (Chadwick et al., 2013). Moreover, using social media can contribute to disabled people gaining positive experiences in terms of friendship and entertainment (Bonilla del Río and Sánchez Calero, 2022), development of social identity and self-esteem, and enjoyment (Caton y Chapman, 2016).

# 3. The Most-Used Social Media Platforms

Social media such as Instagram allows disabled content creators to show their interests, participate in the digital environment and interact with their audience, which helps to reduce barriers and stereotypes and benefits empowerment of this collective (Bonilla del Río et al., 2022b). In fact, some studies conclude that the discourse of Spanish disability-related associations, foundations and public and private organisations on the main social media platforms results in greater knowledge and a positive view of disabled people amongst society as a whole, raising awareness of the stereotypes, prejudices and hate speech that still exist (López Cepeda et al., 2021).

Despite the potential benefits of using social media, the 'digital divide' concept could be applied to individuals who live with a disability (Jaeger, 2012; Werner and Shpigelman, 2019). It is clear that these individuals are less likely than others to have access to computers or the Internet and, therefore, to social media (Hoppestad, 2007). Nevertheless, through accessibility and inclusive design (Pullin, 2011), amongst other matters, the disability 'digital divide' has been successfully bridged in part, as there is still certain digital exclusion that affects individuals depending on their type and degree of disability, their digital skills and their socio-economic circumstances (Dobransky and Hargittai, 2016).

Although social media was designed without disability in mind, it is increasingly used by disabled people, who are more concerned about and interested in using it (Criado et al., 2018). That being said, there are still very few studies on this subject (Enajeh et al., 2018).

The most popular social media platform for disabled people (Asuncion et al., 2012; Raghavendra et al., 2012) and, particularly and specifically, for those with visual impairment (Brinkley y Tabrizi, 2013; Vashistha et al., 2015) is Facebook. This network has been used for free expression, connection with other people and sharing information (Stamou et al., 2016; Smieszek and Borowska-Beszta, 2017; Pritchard, 2021; Aikaterini and Papadopoulos, 2022). Like with other social media, some research has pointed to the empowering effect of Facebook, which is also used as a network that contributes to creating groups and normalising disability (Shpigelman, 2017).

Some other studies were based on analysing use of Instagram by disabled people. This network has also become a platform that provides opportunities for disabled people to defend their and their groups' existence with regard to equality in society (Cahyadi and Setiawan, 2020). Certain studies focus on the visual content posted on Instagram to promote disabled people's abilities and mitigate stereotypes (Mitchell et al., 2021). Other, more general studies analyse, amongst other aspects, self-representation of disability (Cocq and Ljuslinder, 2020; Bonilla del Río et al., 2022b), which contributes to a progressive increase in understanding and acceptance.

The growth undergone by TikTok and the corresponding interest generated in the general public is large and increasingly evident (Schellewald, 2021). However, there are not many studies relating to its use by disabled people and the few existing studies are based on the effectiveness of short videos as a therapy plan (Mohammed, 2022). However, there is also research that aims to raise awareness of the abuse and hurt to which disabled people are sometimes subjected on TikTok (Bonilla del Río, et al. 2022a).

Now, if there is a social media platform that truly contributes to raising awareness and giving a voice to the community with any type of disability, that platform is Twitter (Sarkar et al., 2021). Campaigns raising awareness of disability have caused huge impact on this platform (Gómez Marí et al., 2022). Some studies have focussed on analysing a specific diagnosis (Quadri et al., 2018; Zhang and Ahmed, 2019; Pavlova and Berkers, 2020), whilst others have directed their aims towards studying the content, emotional networks and feelings shared on Twitter relating to disability (Beykikhoshk et al., 2015; Al Zayer and Gunes, 2018; Deriss, 2019; Pal, 2019; Bellon-Harn et al., 2020) or towards debates around disability promoted in the microblogging network (Ineland et al., 2019). Some also expose the daily discrimination and microaggressions faced by disabled people on Twitter (Moral et al., 2022).

## 4. Aims, Hypothesis and Methodology

This study, contextualised on 3rd December 2022, the International Day of Disabled Persons, and in the social network Twitter, precisely aims to analyse the conversation generated on this platform around messages posted with the #disability hashtag on the days prior to and after this specific date. The aim is to describe the predominant participants or organisations and the most prominent topics in the aforementioned conversations. On a day intended to give visibility to individuals with functional diversity and to defend their rights to social inclusion, our starting point is the hypothesis that Twitter would clearly be a setting for institutional messages from organisations related to the cause and for Spanish political leaders supporting inclusiveness. As a second hypothesis, there would be a clear conversational nucleus led by disability activists. The third and last hypothesis is linked to the subject matter, as we can assume there to be assertions around public funding and accessibility, both technological and related to infrastructure.

With the aim of carrying out the study, we examined the configuration of online connections formed around the International Day of Disabled Persons on Twitter, using the social media analysis method (Borgatti et al., 2009; Freeman, 2004). This platform in particular was chosen for its open nature (Williams et al.,2013), which facilitates unlimited observation of information exchanges between users (Wu et al., 2011). 'Twitter offers an interesting environment for research because it generates a huge number of interpersonal interactions that provide a significant set of data' (Pérez-Dasilva et al., 2020, p. 5), allowing the academic research sector to study the information-sharing processes on social media (Brubaker and Wilson, 2018; Benson, 2016). Each and every tweet with the #disability hashtag on Twitter between 27th November and 11th December 2022 were captured; in other words, over the week prior to and the week after the International Day of Disabled Persons.

In total, 18,889 tweets were recorded, containing 17,656 inter-user mentions. These mentions were interpreted as the weighted edges of a graph created with Gephi software, which allowed us to produce the visualisation presented in the results section. Furthermore, users were grouped in hierarchical conglomerates (cluster analysis) (Kaleel and Abhari, 2015) and the network was designed using the Force Atlas 2 brute force algorithm, which places connected nodes close to one another and distances them from unrelated nodes. Using the Louvain algorithm to identify communities, 748 were found, 18 of which were above the 1% node threshold. The partition quality (Modularity) is 0.882. This analysis focusses on the 18 main detected communities (above the 1% node threshold), their topics and the leaders of these conversational nuclei.

## 5. Results

The mentions network obtained over the dates prior to and after the International Day of Persons with Disabilities is enormously dispersed, as can be anticipated and visually evaluated after implementing Force Atlas 2. In fact, the network's main metrics coincide with and reinforce this interpretation of the network's dispersion: Each network node is connected to an average of 1.652 nodes, the average distance between network nodes is 5.695 steps and the maximum distance between the two connected nodes furthest from the network is 19 steps.

Image 1. Graph representing the 18 conversational communities

258 (12,11%)
187 (8,65%)
78 (8,45%)
599 (6,55%)
138 (3,26%)
2 (2,97%)
89 (2,83%)
87 (2,33%)
354 (2,25%)
0 (2,04%)
207 (1,8%)
510 (1,79%)
318 (1,69%)
484 (1,56%)
352 (1,53%)
6 (1,53%)
224 (1,3%)
341 (1,09%)

Source: Prepared by the authors

As can be verified in the methodology, 748 communities were found using the Louvain algorithm. Less than 12% of the network's edges arise between nodes of different communities; therefore, 18 are above the 1% node threshold (Image 1). In other words, in the conversation on the social network Twitter around disability, and over the dates specified in the methodology, 18 conversational communities can be highlighted. 1,618 users participated in the most prominent community (258, in Image 1). The user @Pontifex_es, corresponding to the official account of Pope Francis, received 1,635 mentions, making it the community leader (Image 2). Aside from being the leader of this nucleus, his message is the most re-tweeted in 9 of the 18 analysed communities.

Image 2. The most prominent community, led by Pope Francis

| Comunidad | Num. usuarios |
|---|---|
| 258 | 1618 |
| Total | 1618 |

| Usuario | Menciones |
|---|---|
| Pontifex_es | 1635 |
| vaticannews_es | 36 |
| LaityFamilyLife | 27 |
| SergioSalomonC | 7 |
| FundArcangeles | 5 |
| Catholic_Net | 4 |
| verisimorocha | 4 |
| AztecaChihuahua | 3 |
| CongresoPue | 3 |
| luisfanovich | 3 |

| Tuit | Retuits |
|---|---|
| Hoy queremos recordar a todas las personas con #discapacidad, especialmente a las que sufren porque viven en situación de guerra o se encuentran con una discapacidad a causa de los combates. @laityfamilylife | 2376 |
| RT @Pontifex_es: Hoy queremos recordar a todas las personas con #discapacidad, especialmente a las que sufren porque viven en situación de... | 2376 |
| RT @e_nasarre: Hoy ,3 de Diciembre, es el Día Internacional de las Personas con Discapacidad. No tengo nada que celebrar. Este año ha sido... | 703 |
| Al reunirse con un grupo de personas con #discapacidad, el #PapaFrancisco subrayó que "no hay inclusión si falta la experiencia de la fraternidad y la comunión mutua". "No hay inclusión esta queda como un eslogan". https://t.co/FEhOOIQGaR | 38 |
| RT @vaticannews_es: Al reunirse con un grupo de personas con #discapacidad, el #PapaFrancisco subrayó que "no hay inclusión si falta la exp... | 38 |
| El mensaje de @Pontifex_es para la Jornada Internacional de las Personas con #Discapacidad ha sido publicado también en lengua de señas. Míralo aquí 👇 https://t.co/e1JzusStFG | 16 |
| RT @vaticannews_es: El mensaje de @Pontifex_es para la Jornada Internacional de las Personas con #Discapacidad ha sido publicado también en... | 16 |
| RT @SergioSalomonC: 📅📌 \| Hoy se conmemora el Día Mundial de las Personas con #Discapacidad a fin de promover sus derechos y bienestar en t... | 10 |
| 📅📌 \| Hoy se conmemora el Día Mundial de las Personas con #Discapacidad a fin de promover sus derechos y bienestar en todos los ámbitos de nuestra sociedad. Desde el @CongresoPue, seguimos impulsando una agenda legislativa incluyente, justa y equitativa. #PorPuebla https://t.co/I1WDRanATy | 10 |
| #RT @Pontifex_es: Hoy queremos recordar a todas las personas con #discapacidad, especialmente a las que sufren porque viven en situación de guerra o se encuentran con una discapacidad a causa de los combates. @laityfamilylife | 6 |
| Total | 2376 |

Source: Prepared by the authors

His tweet, a tribute to disabled people, is the most re-tweeted: 'Today we would like to think of everyone with a #disability, especially those living in a state of war (...)' (Image 3). This community covers the general dialogue that arose surrounding International Disability Day between the Pope and the different accounts of Catholic groups or environments, such as @vaticannews_es, a service offered by the Dicastery for Communication of the Holy See, or @Catholic_Net, which defines itself as an online meeting point for Catholics.

Image 3. The most prominent tweet in the most significant community

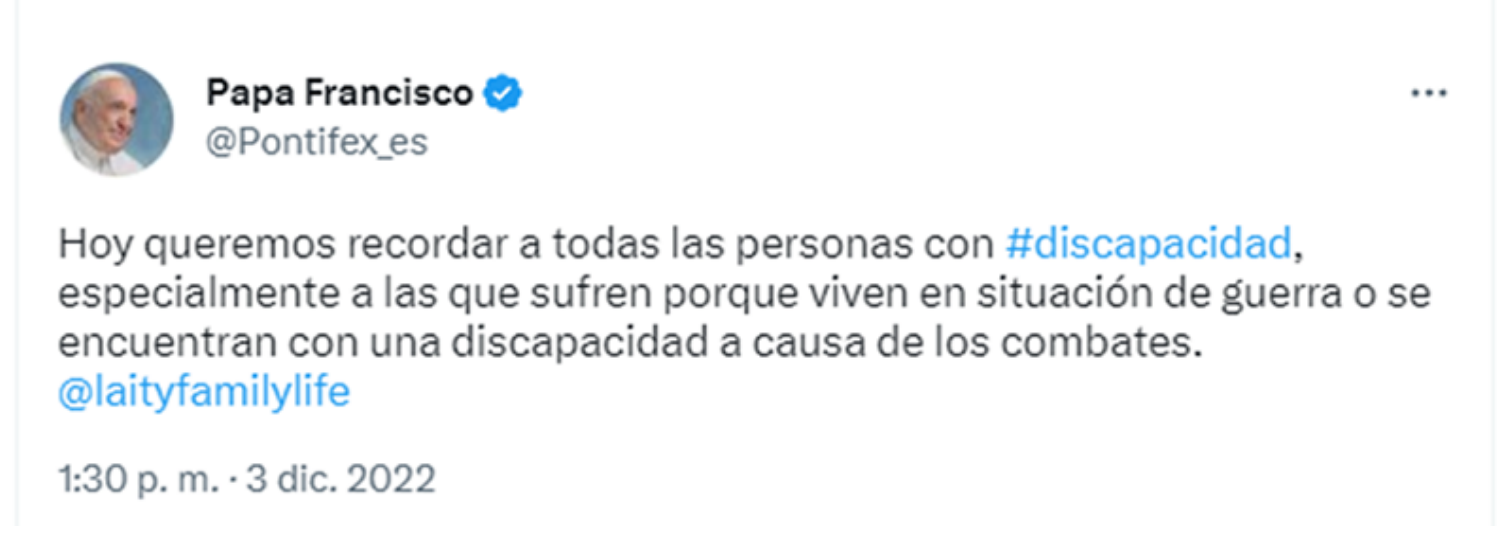

Source: Twitter

1,156 users participated (Image 4) in the second most prominent community (187, Image 1). This is led by the Spanish disability activist Eva Nasarre (@e_nasarre), who was once an icon of aerobics and healthy living in the 80s through her television programme "*Puesta a punto*". Nasarre is now 63 and suffers from a serious degenerative disease that has confined her to a wheelchair. Since she was diagnosed with the disease over twenty years ago, she has spoken out about the injustices experienced by disabled people.

Image 4. The second most prominent community, led by the Spanish activist Eva Nasarre

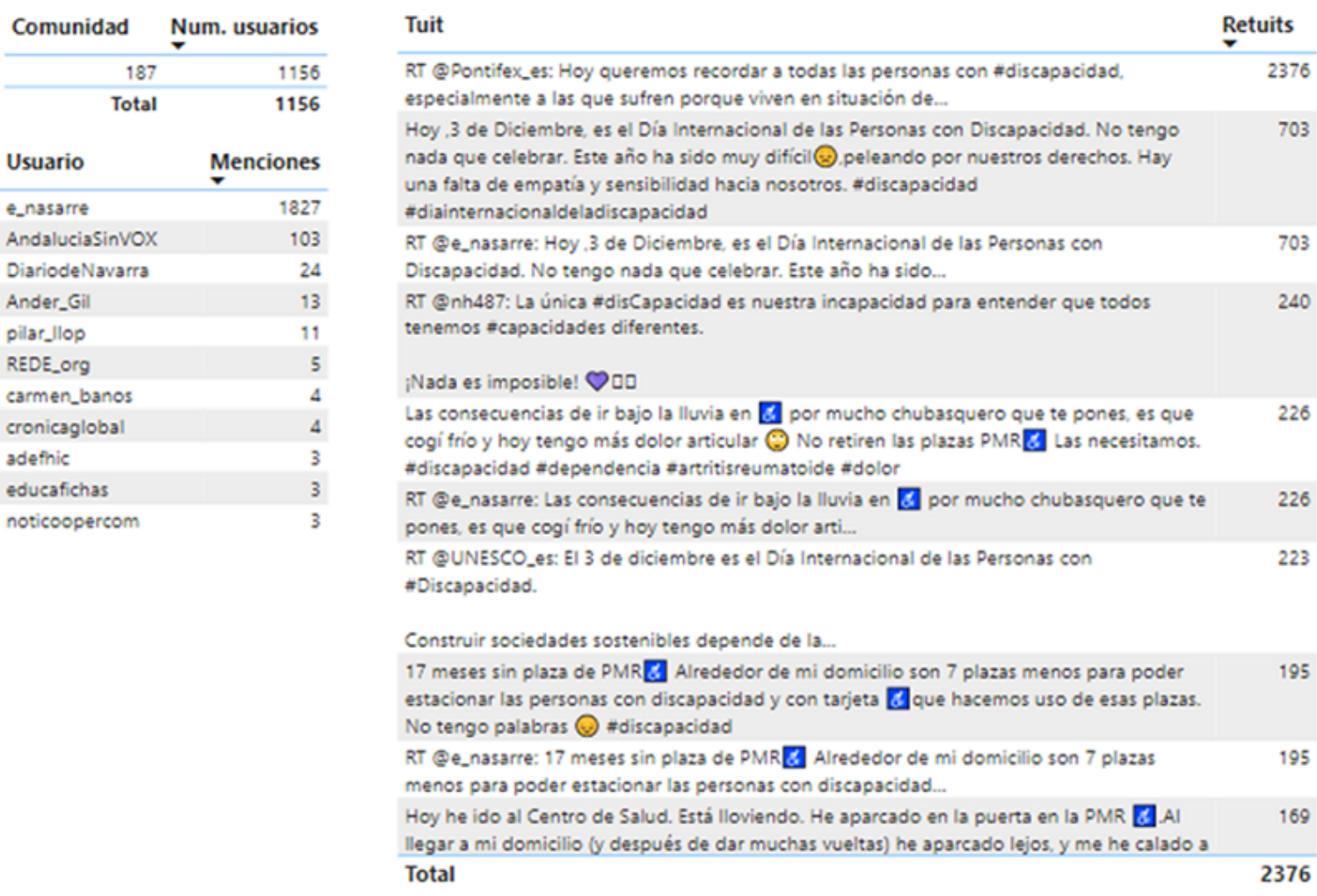

| Comunidad | Num. usuarios |
|---|---|
| 187 | 1156 |
| Total | 1156 |

| Usuario | Menciones |
|---|---|
| e_nasarre | 1827 |
| AndaluciaSinVOX | 103 |
| DiariodeNavarra | 24 |
| Ander_Gil | 13 |
| pilar_llop | 11 |
| REDE_org | 5 |
| carmen_banos | 4 |
| cronicaglobal | 4 |
| adefhic | 3 |
| educafichas | 3 |
| noticoopercom | 3 |

| Tuit | Retuits |
|---|---|
| RT @Pontifex_es: Hoy queremos recordar a todas las personas con #discapacidad, especialmente a las que sufren porque viven en situación de... | 2376 |
| Hoy ,3 de Diciembre, es el Día Internacional de las Personas con Discapacidad. No tengo nada que celebrar. Este año ha sido muy difícil😞,peleando por nuestros derechos. Hay una falta de empatía y sensibilidad hacia nosotros. #discapacidad #diainternacionaldeladiscapacidad | 703 |
| RT @e_nasarre: Hoy ,3 de Diciembre, es el Día Internacional de las Personas con Discapacidad. No tengo nada que celebrar. Este año ha sido... | 703 |
| RT @nh487: La única #disCapacidad es nuestra incapacidad para entender que todos tenemos #capacidades diferentes.<br><br>¡Nada es imposible! 💜 | 240 |
| Las consecuencias de ir bajo la lluvia en ♿ por mucho chubasquero que te pones, es que cogí frío y hoy tengo más dolor articular 🙄 No retiren las plazas PMR♿ Las necesitamos. #discapacidad #dependencia #artritisreumatoide #dolor | 226 |
| RT @e_nasarre: Las consecuencias de ir bajo la lluvia en ♿ por mucho chubasquero que te pones, es que cogí frío y hoy tengo más dolor arti... | 226 |
| RT @UNESCO_es: El 3 de diciembre es el Día Internacional de las Personas con #Discapacidad.<br><br>Construir sociedades sostenibles depende de la... | 223 |
| 17 meses sin plaza de PMR♿ Alrededor de mi domicilio son 7 plazas menos para poder estacionar las personas con discapacidad y con tarjeta ♿que hacemos uso de esas plazas. No tengo palabras 😔 #discapacidad | 195 |
| RT @e_nasarre: 17 meses sin plaza de PMR♿ Alrededor de mi domicilio son 7 plazas menos para poder estacionar las personas con discapacidad... | 195 |
| Hoy he ido al Centro de Salud. Está lloviendo. He aparcado en la puerta en la PMR ♿.Al llegar a mi domicilio (y después de dar muchas vueltas) he aparcado lejos, y me he calado a | 169 |
| Total | 2376 |

Source: Prepared by the authors

Nasarre received 1,827 mentions. Her message - the most re-tweeted after the Pope's - condemns a lack of empathy and sensitivity towards disabled people (Image 5). It is also a prominent message in an additional 5 communities. For many years, she has been an activist for real inclusion of individuals with functional diversity and a staunch defender of the collective's rights. Proof of this are her regular assertions and the profile image she has chosen for her Twitter account: the symbol of disabled people.

Image 5. The most re-tweeted tweet by Eva Nasarre

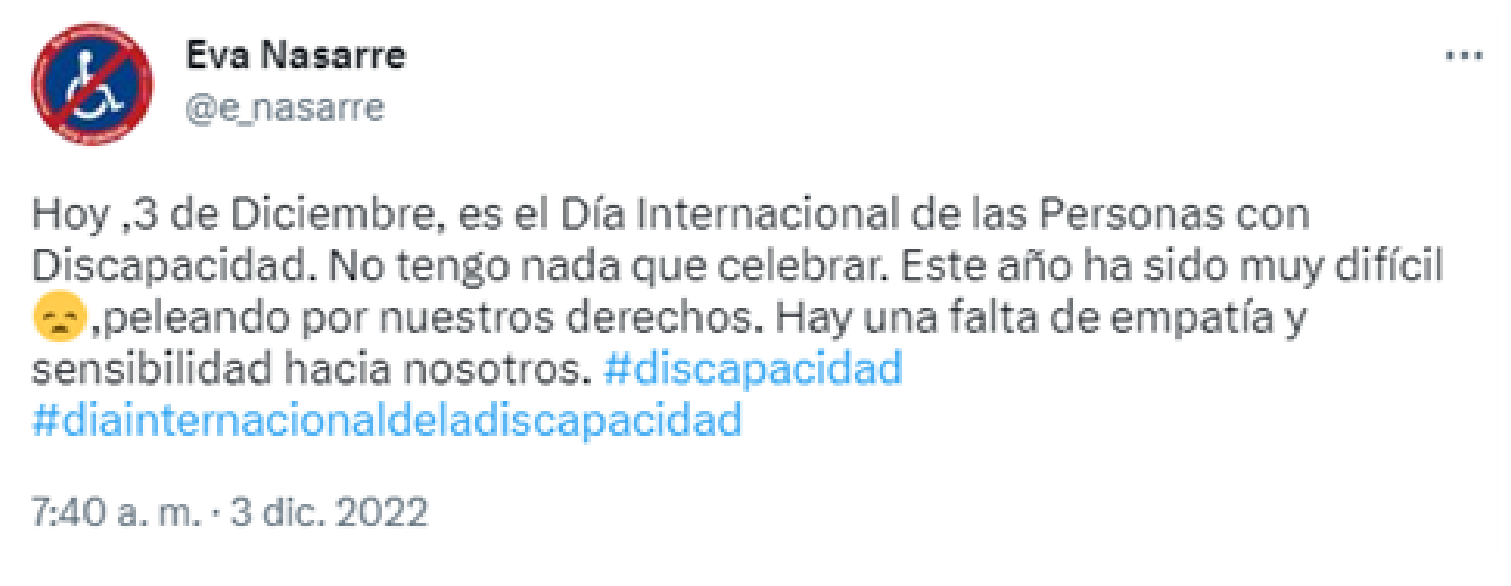

Source: Twitter

The ONCE Foundation (@Fundacion_ONCE) leads the conversation in the third most significant community (78), with 1,129 users. This organisation's account received 245 mentions. However, it is very closely followed by @Plenainclusion (associations defending the rights of people with intellectual or developmental disability and their families), with 217 mentions, and @Cermi_Estatal (Spanish Committee of Representatives of Disabled People), with 200 mentions. In this community, the accounts and users belong to the disability sector and address the topic of violations of disabled people's rights. They are not the only disability-related organisations that lead conversational nuclei. The account @la_discapacidad (Movement for Disability) received 267 mentions and was the leader of another community (138, in Image 1), where the Royal Disability Board (@RPDiscapacidad), with 187 mentions, is also prominent. This is a community with accounts that are closely related to institutions. In the same vein, the account @NoInvisibles ("rare, but not invisible") received 92 mentions, thus leading a conversational nucleus (2, in Image 1) comprising 397 users. This is a community with activists and accounts that are closely related to a specific type of disability or disease, health, law and disabled people's integration into employment. The second most mentioned account in this community is that of Natalia Velilla (@natalia_velilla), a family and disability magistrate. Her tweet, which adds a critical comment, can also be highlighted: "On #DisabilityDay, I'm re-releasing my article on #disability. There's not much more I can say about this topic, which is perhaps one of the most adulterated by supposed do-gooders and which is most suffered in solitude, without true empathy and help from society".

There are Latin American conversational nuclei, such as that led by Carolina Marzán, a Chilean MP and Chair of the Committee of Older People, Disability and Social Development (@carolamarzan). In this community (599), the most mentioned users are Chilean. The most re-tweeted message, after the Pope's tweet, is precisely that posted by the MP herself regarding her dismissal as Chair of the aforementioned Committee (Image 6).

Image 6. Tweet by the Chilean MP Carolina Marzán

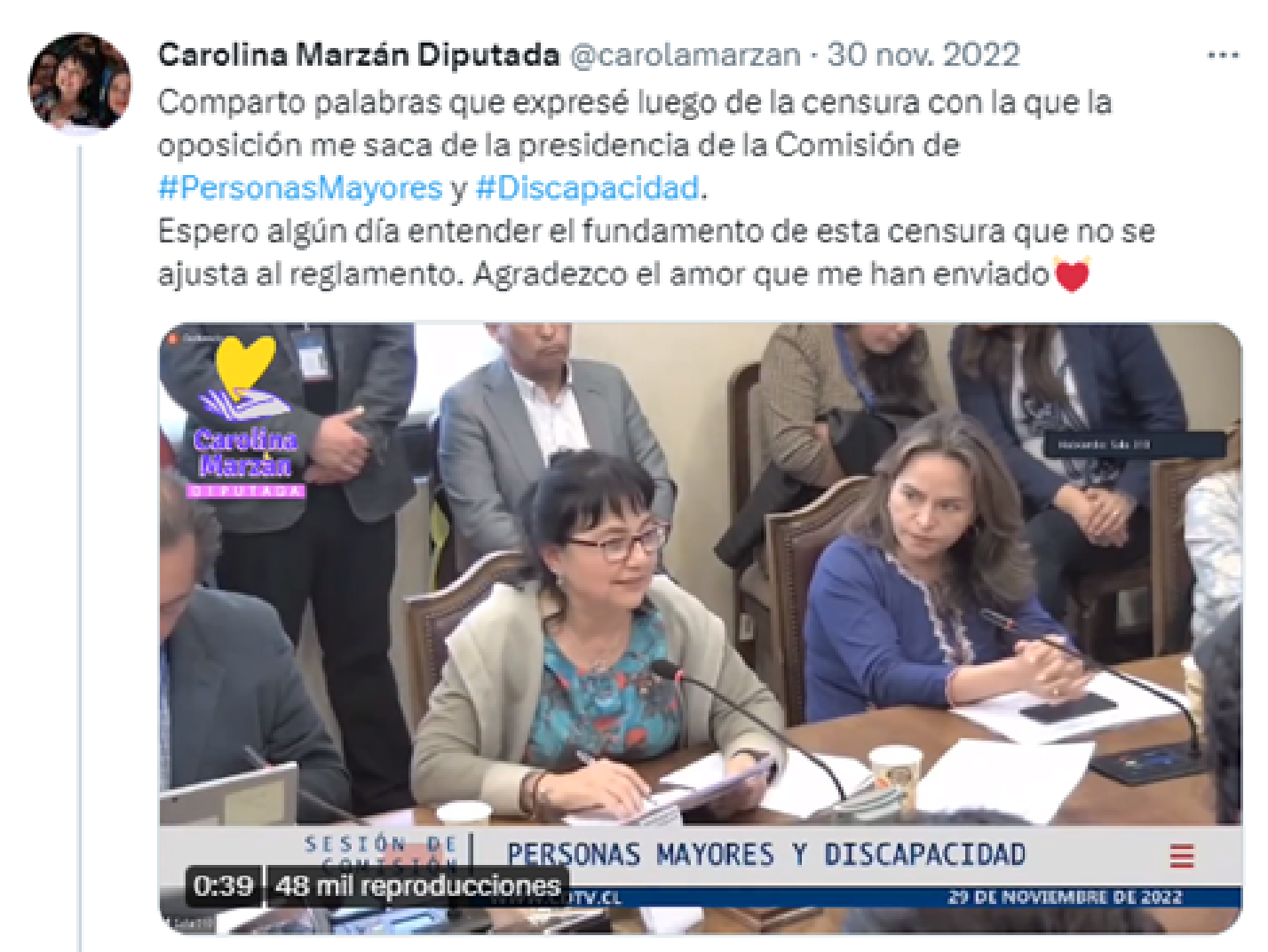

Source: Twitter

The Mexican account @IFT_MX (the official account of the Federal Institute of Telecommunications, the body responsible for regulating, promoting and supervising broadcasting and telecommunications in Mexico) leads a conversation (Community

87) that focusses on employment openings for disabled people and their assertions. The Argentine journalist Lana Montalbán (@LanaMontalban) leads another community (Community 0), where a message by her was the most re-tweeted (Image 7). In this tweet, she criticises a fan who, during the football World Cup, uses a space designated for disabled people although he does not need it.

Image 7. Tweet by the Argentine journalist Lana Montalbán

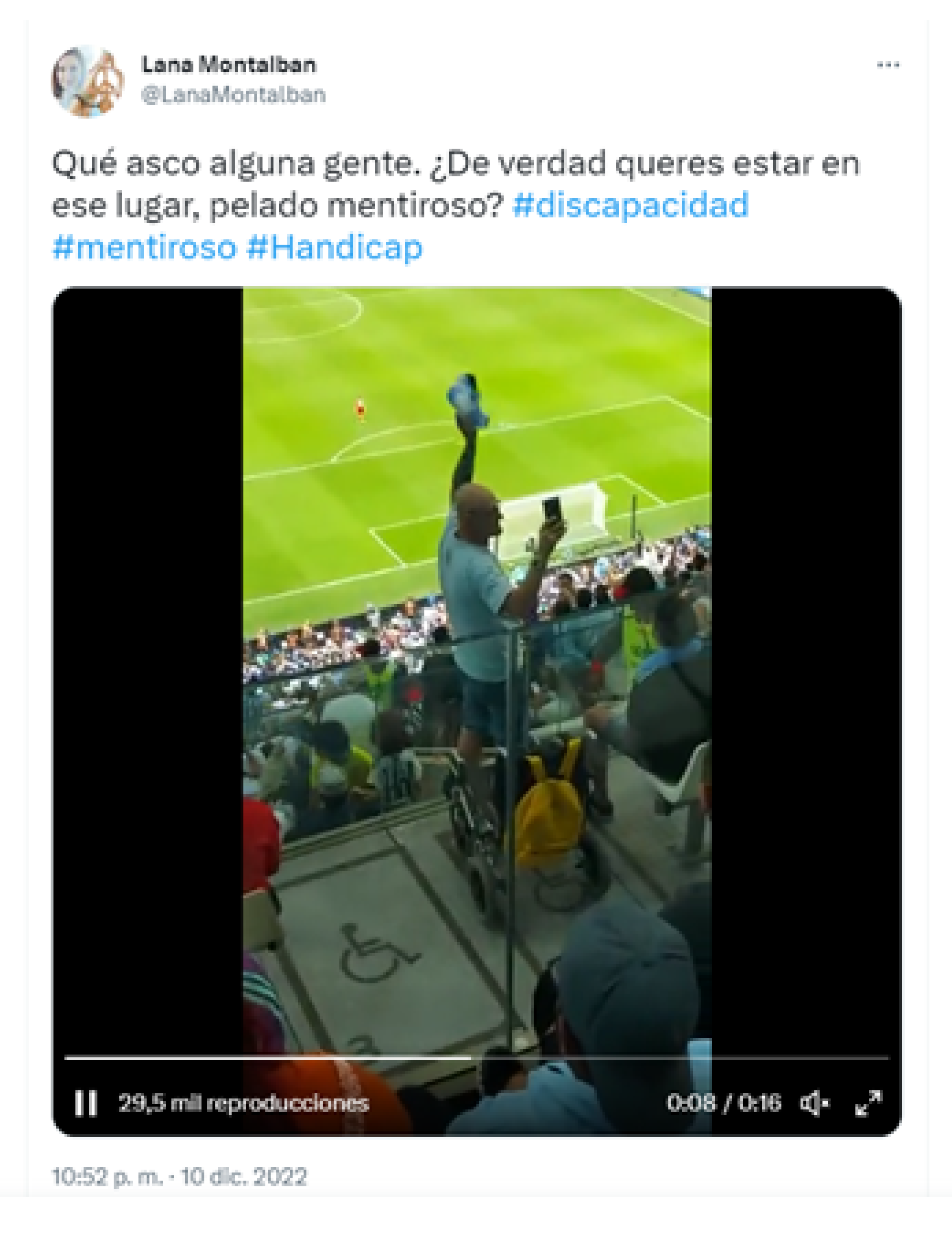

Source: Twitter

Causing less impact - in other words, leading smaller communities - we can highlight the accounts @Supersalud (the official account of the National Health Superintendence, Colombia) and @Disability_SE (María Soledad Cisternas Reyes, SG-NU Special Envoy for Disability and Accessibility, Advocate of Human Rights, Chile). These accounts lead a community (207) with a large presence of Colombian institutional accounts. The Mexican politician Alejandra del Moral (@AlejandraDMV) is the leader of a community (510) with a large presence of Mexican institutional and political accounts. The most re-tweeted message is by Del Moral herself, supporting the integration of disabled people. The conversation revolves specifically around the measures taken by certain Mexican politicians to promote inclusion of disabled people. Another community with a large presence of Latin American users and accounts is that led by the account @UNESCO_es (UNESCO in Spanish), which received 156 mentions (Community 318). This is a nucleus including institutional accounts and those of associations related to disabled people. The most re-tweeted tweet was the Pope's message, as mentioned above. It is very closely followed by, and in the same institutional vein as, that of UNESCO in Spanish (Image 8). The conversation addresses the International Day of Persons with Disabilities and awareness-raising.

Image 8. Tweet marking 3rd December 2022 by UNESCO in Spanish

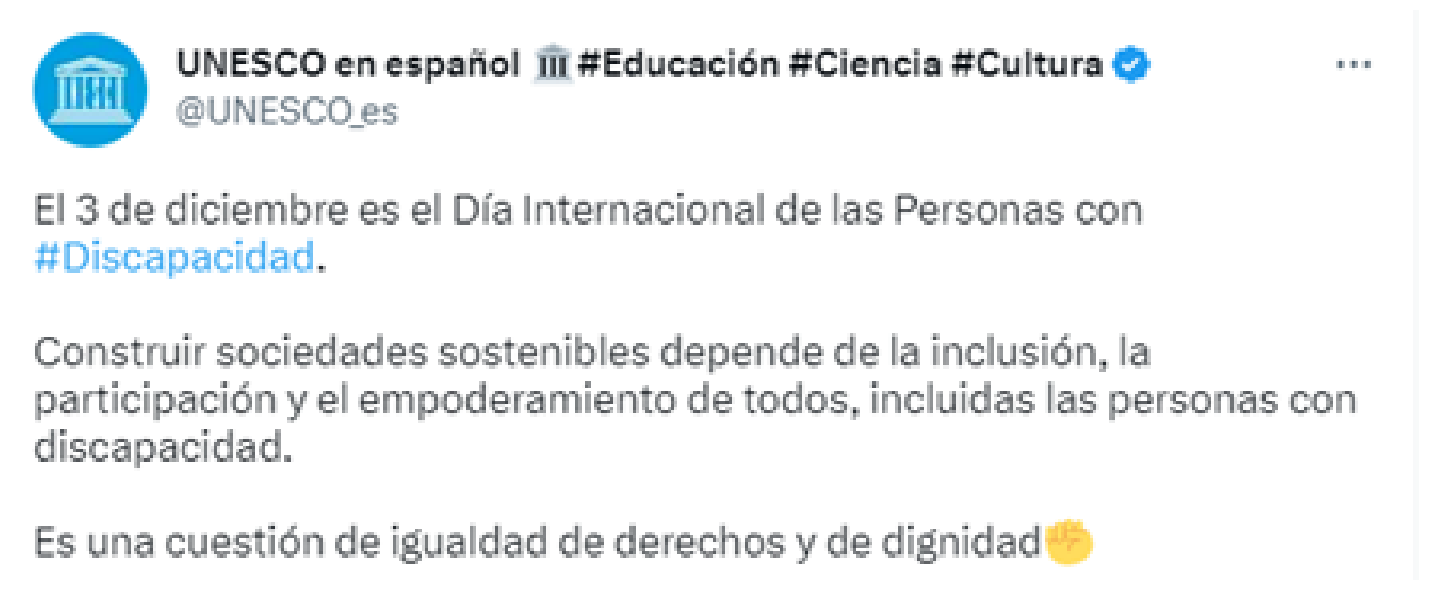

Source: Twitter

The account of the Mexican Government's Secretary of State for Welfare (@bienestarmx) is the leader of a somewhat smaller community (484), which brought together 209 users. Amongst them, there is a large presence of Mexican institutional and political accounts, in addition to a number of associations that work with disabled people. In general, the conversation focusses on different actions designed to share the needs and challenges faced by the different administrations. Although lower in impact, we can find another community (352) in the Latin American environment led by @andiscapacidad (Argentine National Disability Agency) and by the President of Argentina, Alberto Fernández (@alferdez). This nucleus has a large presence of Argentine institutional and political accounts, in addition to a number of associations that work to include disabled people in employment. The two most shared tweets are, on the one hand, that of the Deputy Health Minister of the Government of Buenos Aires Province, Alexia Navarro (@alexiannavarro), announcing an official visit to the events and activities for the International Day of Persons with Disabilities and, on the other, that of the software engineer David Flores (@dfloresdev, an individual committed to web accessibility, a11y, for disabled people) regarding an event organised for 3rd December. The general conversation revolves around events, talks and multiple activities taking place in Argentina on 3rd December.

Also prominent are the communities led by activists focussed on gender perspective, adaptive education and inclusion of the collective of disabled children. For example, Ana Mourelo, who promotes a campaign calling for adapted playgrounds (#1rParcInclusiuBcn) and chairs an association that researches MECP2 duplication syndrome (@AsocMecp2), leads a community (89) of users and accounts related to teaching from a social inclusion perspective. Although the most re-tweeted message in this community is that of Eva Nasarre, Mourelo's awareness-raising message stands out: "Did you know that developing a #disability is like playing the lottery? You think it will never happen to you but the day comes when it does. You don't choose disability, it happens to you. It's up to all of us to make it more visible and to improve disabled people's rights".

The account @acciumred_media (AcciumRed, teachers of easy reading and digital accessibility) leads a community (6) with fewer users. This community is mainly formed of users who are activists with some type of disability, particularly autism. Especially prominent is the message stating: "For anyone with questions about the #murder of a woman with #MultipleSclerosis. 20% of women murdered in #GenderViolence had a #disability. 23% of disabled women have been victims of GV. # 34% of all disabled women are dependent" (Image 9).

Image 9. Tweet with a gender perspective

AcciumRed
@acciumred_media

Para quienes el #asesinato de una mujer con #EsclerosisMúltiple os deja preguntas. El 20% de mujeres asesinadas por #ViolenciaMachista tenía #discapacidad. Un 23% de mujeres con discapacidad ha sido víctima de VM. Del total de mujeres con discapacidad, el 34% es dependiente.

2:15 PM · Dec 9, 2022

Source: Twitter

The general conversation centres around this message and those of other activists and influencers defending the rights of disabled people and awareness-raising of their needs, some of which focus on disabled women or on women caring for disabled children. In other communities, which are smaller but still in the ranking of the 18 most significant, two accounts can be highlighted. On the one hand, the user @paulina 073, who is a mother, carer and activist for the rights and inclusion of disabled people, leads a nucleus of activists who are predominantly deaf, deaf-blind and autistic (Community 224); on the other, we find the account @amanixer (Aragonese association of disabled women), which is mainly followed by institutions in Zaragoza and Aragonese associations. The general conversation covers different plans for the inclusion of disabled people and news items related to the topic, particularly emphasising disabled women as victims of different types of discrimination.

Amongst the 18 communities under analysis, that led by Noah Higón Bellver (@nh487) takes ninth place in terms of impact (354, in Image 1). This 24-year-old Valencian activist has been diagnosed with seven rare diseases, leading her to turn to social media to draw attention to her cause. The community that she leads brings together accounts on rare diseases that aim to inform, raise awareness and defend.

Image 10. Tweet by the Valencian activist Noah Higón

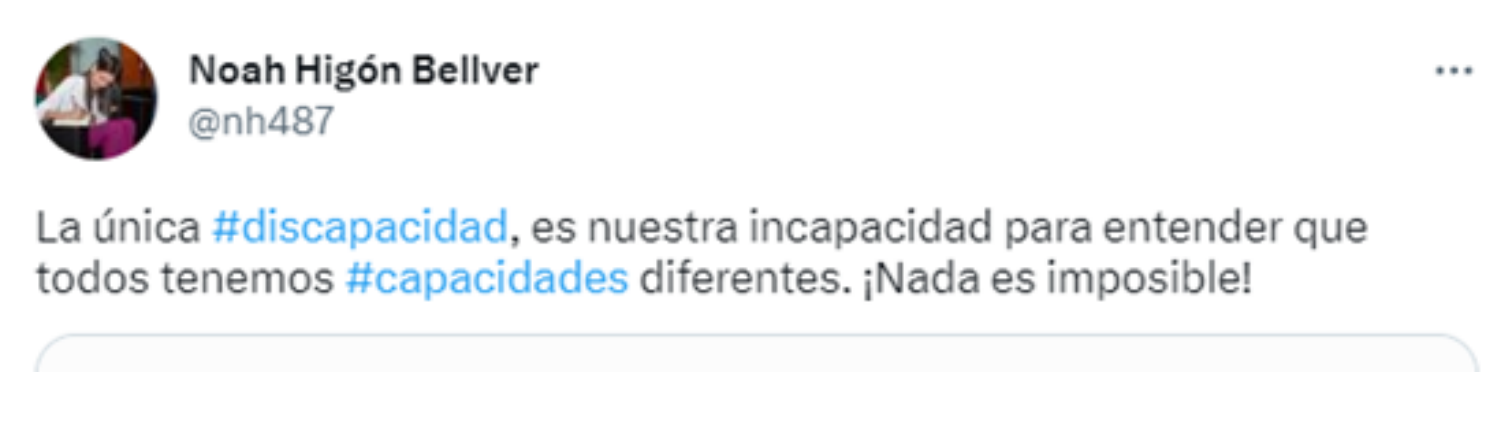

Source: Twitter

In this community, the conversation revolves around the International Day of Persons with Disabilities, and especially prominent are re-tweets of messages by the above-mentioned activists, such as Eva Nasarre, Natalia Velilla and Ana Mourelo.

# 6. Discussion and Conclusions

Our starting point is the idea confirmed by numerous studies on the contribution to activism of the Twitter platform in terms of awareness-raising of and sensitisation to social inclusion issues (Sarkar et al., 2021). Multiple studies have shown that, in this sense, Twitter and other social media play an important role in making disability visible (Foster and Pettinicchio, 2021). However, it remains to be seen if we are close to bridging the 'digital divide' experienced by this collective, as it has fewer opportunities than other individuals to access social platforms (Dobransky and Hargittai, 2016). Another possible subject of discussion is the use made of these networks to achieve real impact of inclusive messages. These platforms may represent an important first step, but awareness-raising and dissemination of integrationist content should not be limited to a single day (Gómez Marí et al., 2022).

After breaking down the conversation generated on this social media platform around the #disability hashtag, and in the context of the International Day of Persons with Disabilities, we can conclude that a general online dialogue was not constructed. Instead, shows of support by Twitter users were made to a few opinion leaders, through re-tweets of their posts. The most prominent community was led by the globally important and popular figure Pope Francis. We can say that this is the only community led by a user not directly related to disability issues and the fight for this collective's rights, beyond his social commitment as a representative who is globally committed to tackling social injustice. In parallel, Spanish institutions and political figures are barely present, in contrast with Latin American networks, where institutional messages are endorsed (h1). In this sense, we can highlight the community led by Carolina Marzán, a Chilean MP and Chair of the Committee of Older People, Disability and Social Development, although her impact is mainly due to her resignation, rather than to awareness-raising of disability. More institutional, and linked to political support, we can see the community constructed around the Argentine president Alberto Fernández, that includes a large presence of institutional accounts, in addition to a number of associations that work to include disabled people in employment. Therefore, there is an absence of prominent Spanish political leaders beyond the indirect presence of Aragonese regional institutions (h1). Nevertheless, and outside any political context, most of the analysed communities are linked to Spanish accounts. This is the case of the community led by the ONCE Foundation, which is mainly formed of users belonging to the disability sector.

We can note a clear predominance of female activists, and Eva Nasarre is the most popular. They are mostly directly-affected individuals or the mothers of disabled children who take on an active role in the fight for awareness-raising, sensitisation and condemnation of violations of disabled people's rights (h2). They sometimes apply a gender focus to these topics, thus speaking out about the double condition of being female and belonging to the disabled collective.

With regard to subject matter (h3), most of the communities revolve around support for general inclusion, but we can highlight criticism linked to a lack of measures and social awareness. There are also references to the 'digital divide'.

All in all, we can say that, in both the institutional and social environments, women are leading the movements and associations that fight to achieve the goals of inclusion and visibility for the disabled collective.

## Authors' contribution

**Terese Mendiguren Galdospin**: Conceptualization, methodology, formal analysis, investigation, writing original draft and visualization. **Koldobika Meso Ayerdi**: Conceptualization, methodology, validation, writing original draft, writing-review and editing, supervision and visualization. **Jesús-ÁngelPérez Dasilva:** Resources, writing-review and editing. **María Ganzabal Learreta**: writing-review and editing, supervision and visualization. **Ainara Larrondo Ureta:** writing-review and editing, supervision and visualization. **Simón Peña Fernández:** writing-review and editing, supervision and visualization. All the authors have read and accepted the published version of the manuscript. Conflicts of interest: The authors declare that they have no conflict of interests.

# Funding

This research was carried out in the framework of the UPV/EHU Project "University, Disability and Inclusion" (US21/23) in partnership with the ONCE Foundation and with the GUREIKER Type A Research Group of the Basque University System (IT1496-22).